\documentclass[sigconf]{acmart}
\copyrightyear{2025}
\acmYear{2025}
\setcopyright{acmlicensed}
\acmConference[SIGIR '25] {Proceedings of the 48th International ACM SIGIR Conference on Research and Development in Information Retrieval}{ July 13--18, 2025}{Padua, Italy.}
\acmBooktitle{Proceedings of the 48th International ACM SIGIR Conference on Research and Development in Information Retrieval (SIGIR '25), July 13--18, 2025, Padua, Italy}
\acmISBN{979-8-4007-1592-1/25/07}
\acmDOI{10.1145/XXXXXX.XXXXXX}
\AtBeginDocument{%
  \providecommand\BibTeX{{%
    Bib\TeX}}}
\usepackage{amsmath,amsfonts}
\usepackage{algorithmic}
\usepackage{array}
\usepackage[caption=false,font=normalsize,labelfont=sf,textfont=sf]{subfig}
\usepackage{textcomp}
\usepackage{stfloats}
\usepackage{url}
\usepackage{cite}
\usepackage{verbatim}
\usepackage{graphicx}
\def\BibTeX{{\rm B\kern-.05em{\sc i\kern-.025em b}\kern-.08em
    T\kern-.1667em\lower.7ex\hbox{E}\kern-.125emX}}
\usepackage{nicematrix}
\usepackage{balance}
\usepackage{multirow}
\usepackage{enumerate}
\usepackage[utf8]{inputenc}
\usepackage{graphicx}
\usepackage{float}
\usepackage{hyperref}
\usepackage{adjustbox}
\usepackage{enumitem}
\usepackage{booktabs}
\usepackage{caption}
\usepackage{xcolor}
\begin{document}

\author{Yu Wang}
\affiliation{%
  \institution{Anhui University}
  \city{Hefei}
  \country{China}}
\email{wangyuahu@stu.ahu.edu.cn}

\author{Lei Sang}
\authornote{Lei Sang is the corresponding author.}
\affiliation{%
  \institution{Anhui University}
  \city{Hefei}
  \country{China}}
\email{sanglei@ahu.edu.cn}

\author{Yi Zhang}
\affiliation{%
  \institution{Anhui University}
  \city{Hefei}
  \country{China}}
\email{zhangyi.ahu@gmail.com}

\author{Yiwen Zhang}
\affiliation{%
  \institution{Anhui University}
  \city{Hefei}
  \country{China}}
\email{zhangyiwen@ahu.edu.cn}

\renewcommand{\shortauthors}{Yu Wang, Lei Sang, Yi Zhang, and Yiwen Zhang.}

\title{Intent Representation Learning with Large Language Model for Recommendation}
\begin{abstract}
Intent-based recommender systems have garnered significant attention for uncovering latent fine-grained preferences. Intents, as underlying factors of interactions, are crucial for improving recommendation interpretability. Most methods define intents as learnable parameters updated alongside interactions. However, existing frameworks often overlook textual information (e.g., user reviews, item descriptions), which is crucial for alleviating the sparsity of interaction intents. Exploring these multimodal intents, especially the inherent differences in representation spaces, poses two key challenges: i) How to align multimodal intents and effectively mitigate noise issues; ii) How to extract and match latent key intents across modalities. 
To tackle these challenges, we propose a model-agnostic framework, \textit{Intent Representation Learning with Large Language Model} (IRLLRec), which leverages large language models (LLMs) to construct multimodal intents and enhance recommendations. Specifically, IRLLRec employs a dual-tower architecture to learn multimodal intent representations. Next, we propose pairwise and translation alignment to eliminate inter-modal differences and enhance robustness against noisy input features. Finally, to better match textual and interaction-based intents, we employ momentum distillation to perform teacher-student learning on fused intent representations.
Empirical evaluations on three datasets show that our IRLLRec framework outperforms baselines\footnote{Code available at \url{https://github.com/wangyu0627/IRLLRec}}. 
%The implementation is available at https://github.com/wangyu0627/IRLLRec.
\end{abstract}
\ccsdesc[500]{Information systems~Recommender systems}
\keywords{Recommendation, Collaborative Filtering, Large Language Models, Intent Modeling, Alignment}
\maketitle
\section{INTRODUCTION}
Recommender systems \citep{2010rs} have become indispensable tools in modern life, enabling personalized content delivery across various domains such as short videos \citep{2020cdprec}, news \citep{2023news}, and E-commerce \citep{zhang2024simplify}. These systems predominantly rely on collaborative filtering (CF) \citep{2017ncf}, which infers user preferences from historical interaction data. Inspired by the advantages of graph neural networks (GNNs \citep{2019ngcf, 2020lightgcn, 2020lrgcn}) in aggregating higher-order collaborative signals, CF models leverage GNNs and graph contrastive learning (GCL \citep{2021sgl, 2022simgcl, 2023lightgcl}) methods to model high-quality representations for users and items. 
Although effective, they largely ignore the latent fine-grained intents between users and items. In real-world scenarios, user-item interactions are often influenced by multiple intent factors \citep{2020DGCF, 2020disenhan}. For instance, users may select skincare products based on specific skin issues like dryness or sensitivity.

Some studies \citep{2019MacridVAE, 2020DGCF, 2022ICLRec} have investigated how intents positively influence recommendation. These works focus on disentangling latent intents from interactions and mapping them to unique feature spaces for modeling. As shown in Figure \ref{fig:challeng} (a) and (b), we introduce multiple intents between users and items to explain how interactions are driven by intent. For instance, the interaction between user $u_2$ and item $i_1$ may be captured by intent $\mathbf{c}_1$, which subsequently recommends more businesses offering food diversity. KGIN \citep{2021kgin} proposes shared intents and leveraged item-side knowledge graphs to capture users' path intents. In contrast, DCCF \citep{2023dccf} enhances self-supervised signals by learning disentangled representations with global context. Although disentangled intents have proven effective in recommendation tasks, the rich semantics conveyed by users in unstructured data are often ignored. In E-commerce, for instance, implicit preferences in user reviews or search logs are challenging to capture solely through explicit interaction data \citep{2024llara}.

The emergence of large language models (LLMs) like GPT-4 \citep{2023gpt4} has recently driven major advancements in user profiling and text summarization for recommendation. Some studies \citep{2024DALR, 2023tallrec, 2024KAR} use the extensive world knowledge of LLMs to deeply describe side information, improving the understanding of complex semantic relationships. For instance, RLMRec \citep{2024rlmrec} focuses on aligning semantic and interaction spaces, while AlphaRec \citep{2024alpharec} enhances recommendation performance by replacing ID-based embeddings with language embeddings. Moreover, using LLMs to model user and item intents not only captures implicit fine-grained preferences but also provides stronger interpretability. However, aligning text-based explicit intents with interaction-based implicit intents remains some significant challenges, as the follow:

\begin{figure*}[t]
    \centering
    \includegraphics[width=0.95\linewidth]{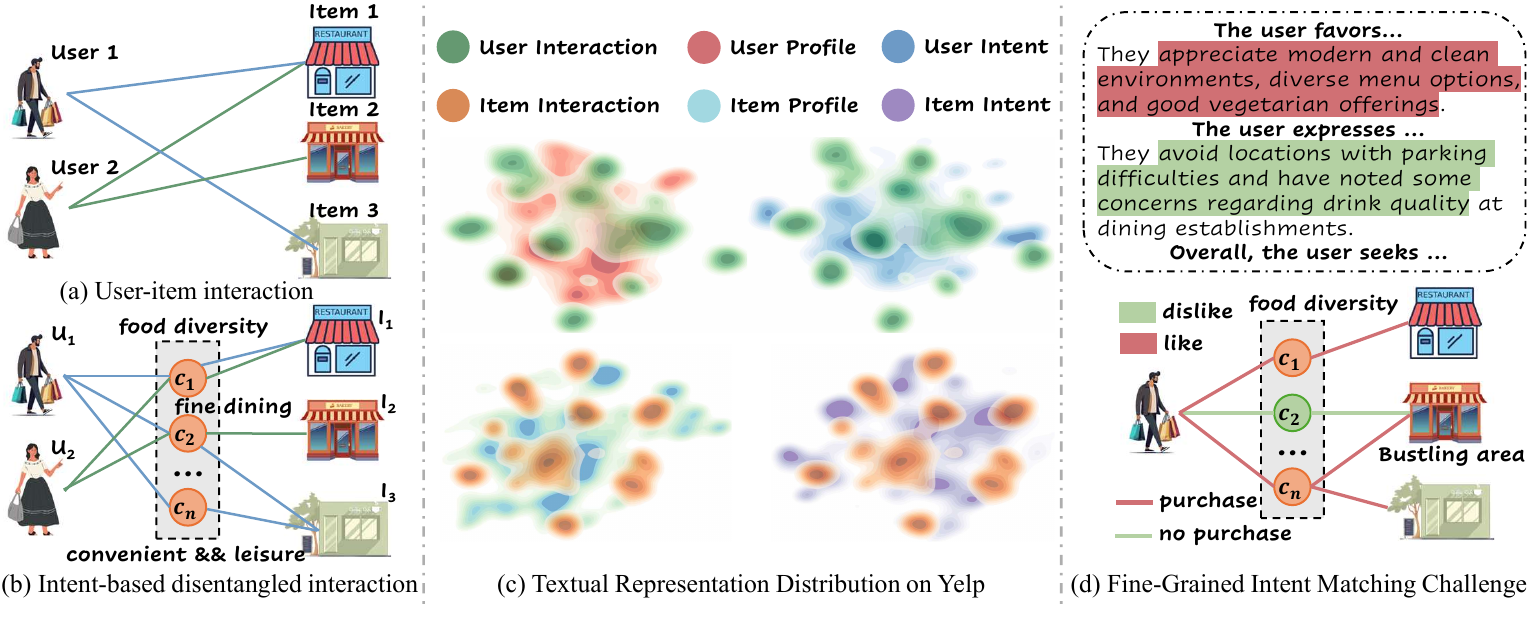}
    \caption{\textnormal{(a) User-item interaction bipartite graph; (b) Disentangled interactions incorporating user intents: $u_1$-$i_3$ is influenced by intents $c_2$ and $c_n$, reflecting a preference for businesses offering fine dining, convenience, and leisure; (c) Gaussian kernel density estimation (KDE \citep{2020kde}) visualizes three embedding types: interaction from the pre-trained LightGCN \citep{2020lightgcn}, profile from RLMRec \citep{2024rlmrec} extracted attribute summaries, and intent from our chain-of-thought reasoning summaries (Figure \ref{fig:user_intent}); (d) The text represents user intents, with red for likes and green for dislikes, and lines indicating interaction or non-interaction.}}
    \label{fig:challeng}
\end{figure*}

\begin{itemize}[leftmargin=*]
    \item \textbf{Cross-modal semantic alignment.}
    Semantic alignment can bridge the gap between unstructured textual data and structured user-item interactions, effectively capturing fine-grained intents in both modalities to enhance contextual representations of users and items. However, the inherent differences between textual and interaction data in the representation space require complex mapping mechanisms \citep{2024alpharec} to ensure consistency and semantic fidelity. 
    As illustrated in Figure \ref{fig:challeng} (c), profile and intent representations overlap with interaction-based representations but differ significantly in distribution. This suggests that intents capture behavioral drivers often missed by fixed profiles in interactions. Moreover, during the alignment process, misaligned representations influenced by noise or sparse interaction data \citep{2024rlmrec} can lead to suboptimal intent fusion.
    \item \textbf{Fine-Grained Intent Matching.}
    In multimodal data, key information is often embedded in specific text or interactions. Precise matching mechanisms \citep{2021ALBEF} extract relevant needs and behaviors from fragmented signals, capturing fine-grained intents. For example, as shown in Figure \ref{fig:challeng} (d), a user may express likes (e.g., "diverse menu options") alongside dislikes (e.g., "parking difficulties"). These likes align with interaction intents like $\mathbf{c}_1$'s food diversity, while dislikes correspond to $\mathbf{c}_2$'s bustling area.
    Additionally, redundant textual information (e.g., general environment descriptions \citep{2024DHGPF}) and noisy interaction signals (e.g., misclicks or popularity bias \citep{2024GBSR}) reduce matching accuracy.
\end{itemize}

To tackle the challenges mentioned above, we propose an Intent Representation Learning with Large Language Model (\textbf{IRLLRec}) framework for recommendation. 
First, for \textbf{Challenge 1}, a dual-tower-based \textbf{Intent Alignment} (IA) module is proposed, aiming to bring intents in two distinct representation spaces (text and interaction) closer and to capture fine-grained semantics advantageous for recommendations. IA employs specific encoders to map multimodal intents into a shared space for fusion. It includes two alignment strategies: pairwise alignment enhances consistency by maximizing multimodal mutual information, to mitigate spatial differences. Translation alignment perturbs multimodal representations of individual users and items, improving robustness to input noise features.
For \textbf{Challenge 2}, we design a \textbf{Interaction-text Matching} (ITM) module to extract and match latent key intents for precise preference recommendations. ITM employs momentum distillation \citep{2021ALBEF} to perform teacher-student learning on the fusing intent representations. It also maps the dual-tower encoders into two momentum unimodal encoders to generate pseudo-labels for the teacher model, aiming to identify the optimal matching positions for multimodal intents.
Overall, IRLLRec is an LLM-based model-agnostic recommendation framework that considers the potential of multimodal intents to enhance traditional recommendation models and effectively mitigating noise and matching issues in intent alignment. The contributions are summarized as follows:

\begin{itemize}[leftmargin=*]
\item We investigate multimodal intent extraction methods and introduce IRLLRec, a dual-tower alignment paradigm that effectively captures fine-grained semantic information for recommendations. This framework utilizes mutual information maximization and noise perturbation to generate high-quality representations.
\item We further propose an Interaction-Text Matching module, which employs momentum distillation for teacher-student learning on fusing intent representations. This approach accurately extracts latent key intents and enables cross-modal matching.
\item We conducte extensive experiments on three public datasets, integrating state-of-the-art recommendation methods. IRLLRec outperformed existing baselines and significantly enhanced the base models. Furthermore, we analyzed how intent alignment and matching contribute to optimizing recommendation outcomes.
\end{itemize}

\section{PRELIMINARIES}
\noindent{\textbf{Collaborative Filtering.}}
In a general recommendation model, there exists a set of $M$ users and a set of $N$ items, represented as $\mathcal{U} = \{u_1, u_2, \ldots, u_M\}$ and $\mathcal{I} = \{i_1, i_2, \ldots, i_N \}$, respectively. According to previous research \citep{2020lightgcn, 2022simgcl}, historical interactions between these users and items are stored in a user-item interaction matrix $\mathbf{R} \in \mathbb{R}^{M \times N}$. The edges, denoted as $\mathcal{E}{ui} = \mathcal{E}{iu} = R_{ui}$, correspond to the entries in $\mathcal{G} = \langle \mathcal{V} = {\mathcal{U}, \mathcal{I}}, \mathcal{E} \rangle$. This formulation frames the recommendation problem as a bilateral graph generation task. The probability of the existence of edges in the user-item interaction graph $\mathcal{G}$ is calculated using the following formula:
\begin{equation}
    \mathbb{P}(\hat{R}_{ui} = 1 | \mathbf{e}_u, \mathbf{e}_i) = \sigma(\mathbf{e}_u^\top \mathbf{e}_i),
    \label{eq1}
\end{equation}
where $\sigma$ is the sigmoid function. $\mathbf{e}_u$ and $\mathbf{e}_i$ are the embeddings of user $u$ and item $i$ via encoding, respectively.

\noindent{\textbf{Text-enhanced Recommendation.}}
Recent works \citep{2024rlmrec, 2024alpharec} draw inspiration from the powerful summarization capabilities of LLM, leveraging rich textual information such as user reviews and item descriptions in interactions to enhance collaborative filtering. Existing study \citep{2022CoT} indicate that system prompts can effectively alleviate LLM hallucinations and improve the quality of its responses. Therefore, the common approach is to design specific prompts $S_{u/v}$ for each user $u$ and item $i$ to obtain summarization responses or profiles from the LLM. This process is represented as follows:

\begin{equation}
    \mathcal{P}_u = \text{LLMs}(S_u, Q_u), \quad \mathcal{P}_i = \text{LLMs}(S_i, Q_i),
    \label{eq2}
\end{equation}
where $Q_{u/i}$ represents user/item profile generation prompts, and $\mathcal{P}_{u/i}$ serves as input for the subsequent stage. Formally, the semantic information about users and items is transformed into features $\mathbf{s}_{u/v} = \text{LLMs}_{emb} (\mathcal{P}_{u/v})$ by the LLM's transformer \citep{2017transformer} and integrated into the recommender system.

\noindent{\textbf{InfoNCE-based Alignment.}}
In recent years, contrastive learning (CL \citep{2020simclr}) has demonstrated significant advantages in cross-modal alignment by bringing the positive samples of two modalities closer in embedding space while pushing the negative samples apart. The core component of CL, the InfoNCE loss \citep{2020simclr}, has been proven to be a critical means of Mutual Information Maximization (MIM). To mitigate the interference of irrelevant information (noise) on recommendations, some studies \citep{2024rlmrec, 2024DALR} typically use it to align two modalities. Formally,

\begin{equation}
    \mathcal{L}_{\text{InfoNCE}} = \mathbb{E}_{p(\mathbf{e}, \mathbf{s})} \left[ f_{\text{sim}}(\mathbf{e}, \mathbf{s}) - \log \sum_{\mathbf{s}' \in \mathcal{B}} \exp f_{\text{sim}}(\mathbf{e}, \mathbf{s}') \right].
    \label{eq3}
\end{equation}
where $f_{\text{sim}}$ represents the similarity function, usually the inner product or cosine similarity, and $\mathbf{e}$, $\mathbf{s}$, $\mathbf{s}'$ are the embeddings of users, positive samples, and negative samples in a batch $\mathcal{B}$. The loss optimization guides the similarity of positive sample pairs to be higher than that of negative sample pairs.

\begin{figure*}[t]
    \centering
    \includegraphics[width=\linewidth]{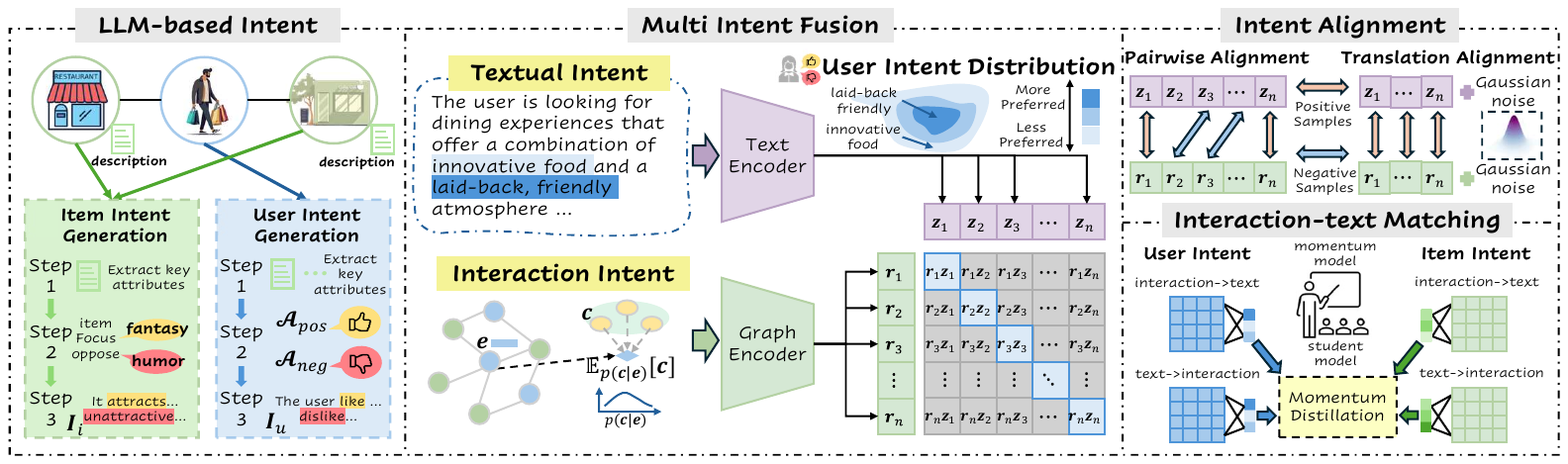}
    \caption{Illustration of IRLLRec. Multi Intent Fusion (MIF): \textnormal{MIF takes textual and interaction-based intents as inputs, learning intent embeddings $\mathbf{z}$ and $\mathbf{r}$ through a dual-tower model and fusing them.} Intent Alignment (IA): \textnormal{IA bridges spatial discrepancies by aligning two distinct representation spaces.} Interaction-text Matching (ITM): \textnormal{ITM employs momentum distillation for teacher-student learning, enabling optimal matching of multimodal intents for users and items.}}
    \label{fig:model}
\end{figure*}

\section{METHODOLOGY}
In this section, we propose an intent-enhanced framework, IRLLRec, designed to assist recommendation models capture the fine-grained preferences of users and items hidden in interactions. As shown in the figure \ref{fig:model}, we will introduce the model's inputs and the three key modules of IRLLRec step by step.

\subsection{LLM-based Intent Construction}
\textbf{Textual intent.} As illustrated in the Figure \ref{fig:model}, constructing textual intent follows a structured multi-step process designed to extract and summarize user and item preferences, encoding them into fine-grained intent representations to improve recommendation. 

\begin{figure}[h!tp]
    \centering
    \includegraphics[width=\linewidth]{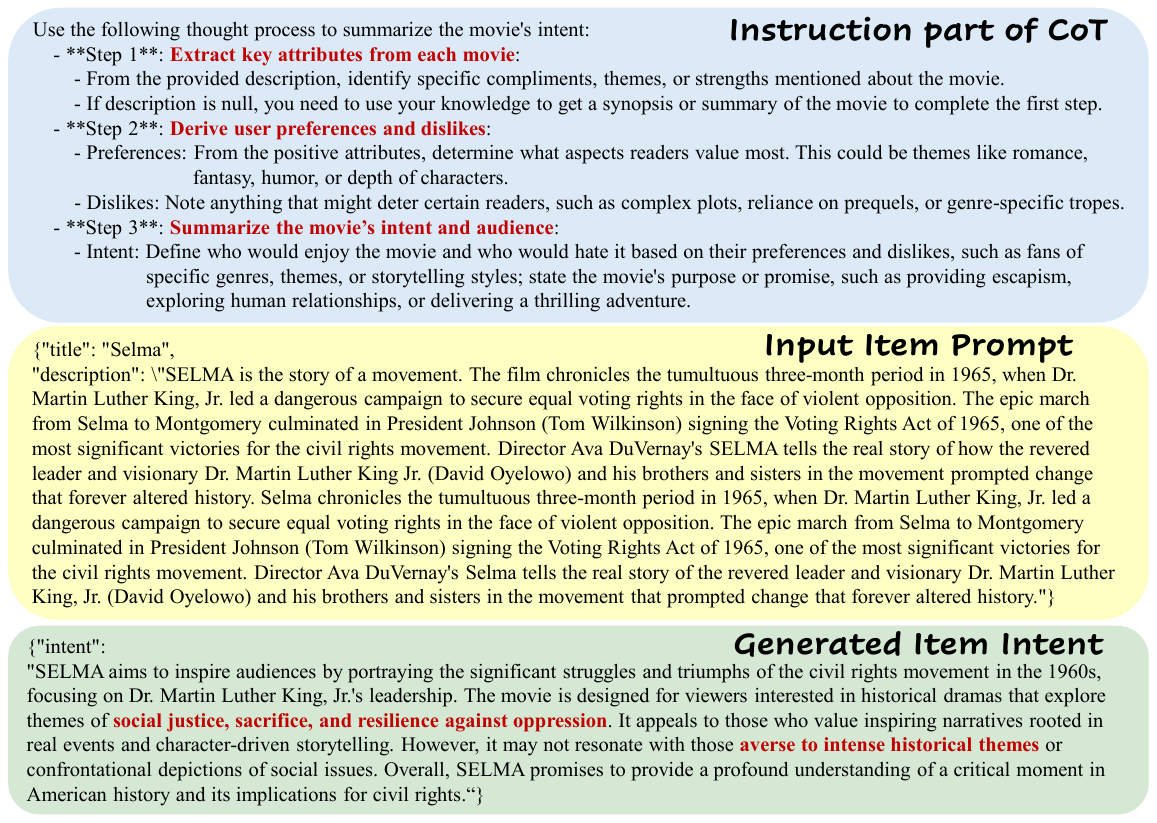}
    \caption{Case study on LLM-based item intent generation in Amazon-movie dataset.}
    \label{fig:item_intent}
\end{figure}

\begin{figure}[h!tp]
    \centering
    \includegraphics[width=\linewidth]{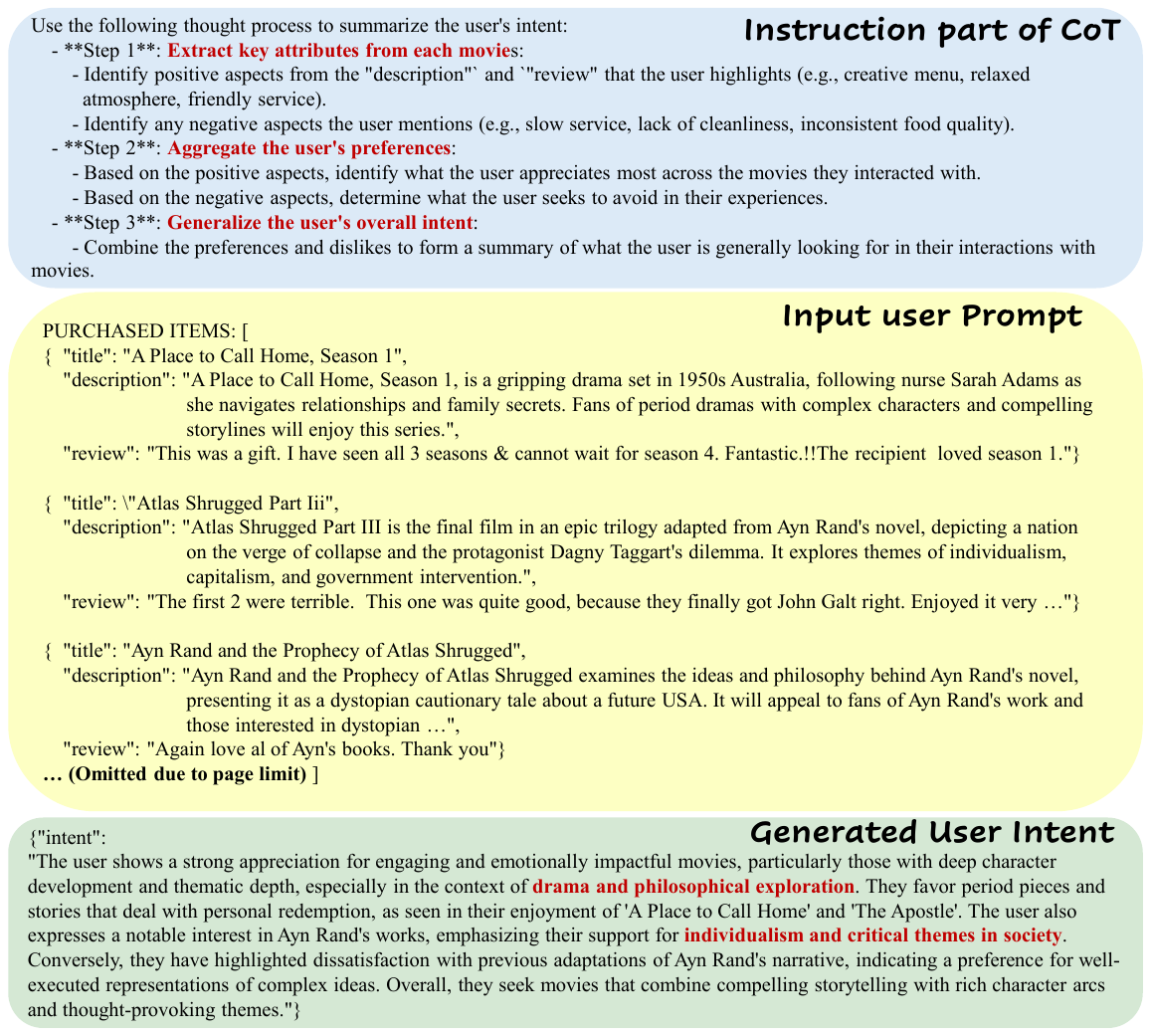}
    \caption{Case study on LLM-based user intent generation in Amazon-movie dataset.} 
    \label{fig:user_intent}
\end{figure}

The first step involves summarizing the intents, where Chain of Thought (CoT \citep{2022CoT}) prompts are crafted to incrementally direct the LLM to produce well-defined preferences, ensuring a diverse range of intents. As shown in Figures \ref{fig:item_intent} and \ref{fig:user_intent}, we summarize user and item intents as $I_{u/i} = \text{Aggregate}(\{\mathcal{A}_{\text{pos}}, \mathcal{A}_{\text{neg}}\})$ where $\mathcal{A}_{\text{pos}}$ and $\mathcal{A}_{\text{neg}}$ represent the user's or item's likes and dislikes. The LLM extracts features of each item to obtain $I_{i}$, while $I_{u}$ analyzes the features of items in each interaction to deeply explore the user's intents. The specific forms are as follows:
\begin{equation}
    I_{u} = \text{Aggregate} \left( \bigcup_{i \in \mathcal{I}_{u}} \Big\{ \mathcal{A}_{\text{pos}}(i), \mathcal{A}_{\text{neg}}(i) \Big\} \right).
    \label{eq4}
\end{equation}
where $\mathcal{I}_{u}$ represents the set of items interacted with by user $u$. "Aggregate" refers to the third step in the CoT prompt, summarizing the extracted information to obtain the final $I_{u}$ and $I_{i}$. Subsequently, following \citep{2024rlmrec}, the textual descriptions of these intents are fed into a decoder-only LLM to generate the intent features $\mathbf{x}_u = \text{LLMs}_{emb} (I_{u})$ and $\mathbf{x}_i = \text{LLMs}_{emb} (I_{i})$ for users and items.

\noindent \textbf{Interaction intent.}
Finally, we model the interaction-based intents $\mathbf{c}_u$ and $\mathbf{c}_i$ for users and items. User intent refers to the user's needs in a specific context (e.g., preferred brands or movie genres), while item intent refers to the item's context, such as products with a "health" label being suitable for fitness enthusiasts. In this study, we assume that the latent intents $\mathbf{c}_u$ and $\mathbf{c}_i$ follow the distributions $P(\mathbf{c}_u \mid u)$ and $P(\mathbf{c}_i \mid i)$, respectively. According to statistical derivation, when the interaction probability $P(y \mid \mathbf{c}_u, \mathbf{c}_i)$ between users and items is modeled as the joint distribution of latent intent factors, this probability can be expressed as the following:
\begin{equation}
    P(y \mid u, i) = \mathbb{E}_{P(\mathbf{c}_u \mid u) P(\mathbf{c}_i \mid i)} \left[ P(y \mid \mathbf{c}_u, \mathbf{c}_i) \right],
    \label{eq5}
\end{equation}
where $P(\mathbf{c}_u \mid u)$ and $P(\mathbf{c}_i \mid i)$ describe the intent distributions of users and items, respectively. Furthermore, to simplify the computation, the mathematical expectation of the prediction function $f(\cdot)$ can be approximated as follows:
\begin{equation}
    \mathbb{E}_{P(\mathbf{c}_u \mid u) P(\mathbf{c}_i \mid i)} \left[ f(\mathbf{c}_u, \mathbf{c}_i) \right] \approx f \left( \mathbb{E}_{P(\mathbf{c}_u \mid u)}[\mathbf{c}_u], \mathbb{E}_{P(\mathbf{c}_i \mid i)}[\mathbf{c}_i] \right).
    \label{eq6}
\end{equation}
where $\mathbb{E}_{P(\mathbf{c}_u \mid u)}[\mathbf{c}_u]$ and $\mathbb{E}_{P(\mathbf{c}_i \mid i)}[\mathbf{c}_i]$ represent the expectations of $\mathbf{c}_u$ and $\mathbf{c}_i$, respectively. The above formulas provide a theoretical basis for modeling the interaction probability between users and items in recommender systems. Specific mathematical derivations can be found in the references \citep{2021DeRec,2023dccf}.

\subsection{Multi Intent Fusion}
In this section, we configure two encoders: a text encoder and a graph encoder. Inspired by \citep{2024alpharec} and \citep{2024DALR}, we designe the text encoder as a linear mapping, which effectively controls computational cost and aligns the textual space with the interaction space. The specific formula is as follows:
\begin{equation}
    \mathbf{z} = \mathbf{W}_2 (\sigma (\mathbf{W}_1 \mathbf{x} + \mathbf{b}_1)) + \mathbf{b}_2,
    \label{eq7}
\end{equation}
where $\sigma$ is the LeakyReLU function, and $\mathbf{W}$ and $\mathbf{b}$ are the weight matrix and bias vector, corresponding to the linear transformation and bias term, respectively. This mapping has been proven effective in capturing fine-grained preferences in the textual space, laying the foundation for multi-intent alignment. Next, we empirically adopt LightGCN \citep{2020lightgcn} as the graph encoder because the message-passing mechanism of graphs is widely recognized for its advantages in capturing collaborative signals. The representations $\mathbf{E}_{u}$ and $\mathbf{E}_{i}$ for users and items are shared parameters in most recommenders, and our update mechanism is as follows:
\begin{equation}
    \mathbf{e}_u^{(l)} = \sum_{i \in \mathcal{N}_u} \frac{1}{\sqrt{|\mathcal{N}_u||\mathcal{N}_i|}} \mathbf{e}_i^{(l-1)}, \mathbf{e}_i^{(l)} = \sum_{u \in \mathcal{N}_i} \frac{1}{\sqrt{|\mathcal{N}_u||\mathcal{N}_i|}} \mathbf{e}_u^{(l-1)},
    \label{eq8} 
\end{equation}
where $l$ is the encoder layer, and $\mathbf{e}_u^{(0)}$ and $\mathbf{e}_i^{(0)}$ denote one of the nodes $\mathbf{E}_{u}$ and $\mathbf{E}_{i}$. $\mathcal{N}_u$ and $\mathcal{N}_i$ respectively represent the first-order receptive fields of users and items.

However, the intents of users and projects are often diverse, making it challenging to fully capture potential preferences with a global representation alone. To address this, we further introduce the concept of interaction intent modeling. By employing the mechanism defined in Eq. \ref{eq6} and incorporating $K$ intent prototypes, the global representations $\mathbf{e}_u$ and $\mathbf{e}_i$ are mapped to intent-aware embeddings $\mathbf{r}_u$ and $\mathbf{r}_i$, as shown below:
\begin{equation}
\begin{aligned}
    \mathbf{r}_u^{(l)} &= \mathbb{E}_{P(\mathbf{c}_u \mid \mathbf{e}_u^{(l)})} \left[ \mathbf{c}_u \right] = \sum\nolimits_{k=1}^{K} \mathbf{c}_u^k P(\mathbf{c}_u^k \mid \mathbf{e}_u^{(l)}), \\
    \mathbf{r}_i^{(l)} &= \mathbb{E}_{P(\mathbf{c}_i \mid \mathbf{e}_i^{(l)})} \left[ \mathbf{c}_i \right] = \sum\nolimits_{k=1}^{K} \mathbf{c}_i^k P(\mathbf{c}_i^k \mid \mathbf{e}_i^{(l)}),
    \label{eq9}
\end{aligned}
\end{equation}
where $\mathbf{r}_u^{(l)}$ represents the intent-aware representation of user $u$ at layer $l$. The $\mathbf{c}_u^k$ and $\mathbf{c}_i^k$ are the intent prototypes of the user and item, respectively. To obtain these probability distributions, we need to further clarify the relevance between each user or item and the intent prototypes. The $\mathbb{E}$ of $\mathbf{c}$ is defined as follows:
\begin{equation}
\begin{aligned}
    P(\mathbf{c}_u^k \mid \mathbf{e}_u^{(l)}) &= {\exp (\mathbf{e}_u^{(l)} \mathbf{c}_u^k)} / {\sum\nolimits_{k'=1}^{K} \exp (\mathbf{e}_u^{(l)} \mathbf{c}_u^{k'})}, \\
    P(\mathbf{c}_i^k \mid \mathbf{e}_i^{(l)}) &= {\exp (\mathbf{e}_i^{(l)} \mathbf{c}_i^k)} / {\sum\nolimits_{k'=1}^{K} \exp (\mathbf{e}_i^{(l)} \mathbf{c}_i^{k'})},
    \label{eq10}
\end{aligned}
\end{equation}

Considering that intents are susceptible to noise in interactions, we employ a graph structure learning (GSL \citep{sang2024intent, 2021PTDNet}) approach to reconstruct a clean interaction graph and filter intent representations. GSL has been widely applied to various graph tasks and proven effective in alleviating noise issues, summarized as follows: $\mathcal{G}^{(l)} = \mathbf{M}^{(l)} \odot \mathcal{G}$, where the mask matrix $\mathbf{M}_{ij}^{(l)}$ assigns a weight to each edge in $\mathcal{G}$. The closer the value is to 0, the less significant the interaction, and vice versa. The formula is as follows:
\begin{equation}
    \mathbf{M}^{(l)}_{ij} = \mathbf{D}^{-1} \cdot \alpha_{ij}, \quad \alpha_{ij} = (s(\mathbf{r}_u^{(l)}, \mathbf{r}_i^{(l)}) + 1)/{2}
    \label{eq11}
\end{equation}
where $s(,)$ is the cosine similarity, and normalization is used to obtain the weight $\alpha_{ij}$ within the range $[0,1]$. $\mathbf{D}$ is the degree matrix of the graph $\mathcal{G}^{(l)}$. This method emphasizes the importance of low-degree nodes in the graph (through inverse degree weighting), thereby preventing high-degree nodes (e.g., hub nodes) from excessively dominating the edge weight distribution. 
\begin{equation}
    \mathbf{R}_u^{(l)} = \mathcal{G}^{(l)} \cdot \mathbf{R}_u^{(l)}, \quad \mathbf{R}_i^{(l)} = \mathcal{G}^{(l)} \cdot \mathbf{R}_i^{(l)}.
    \label{eq12}
\end{equation}
Finally, we perform an averaging operation on the intent representations to obtain $\mathbf{R}_u = {1}/{L} \sum_{l=1}^L \mathbf{R}_u^{(l)}$ and $\mathbf{R}_i = {1}/{L} \sum_{l=1}^L \mathbf{R}_i^{(l)}$.

\subsection{Intent Alignment}
This section is divided into two alignment strategies: pairwise alignment and translation alignment. As illustrated in Figure \ref{fig:model} under ``Intent Alignment", pairwise alignment aims to align the two modalities using interaction-based CF information, while translation achieves the same goal through simple sample shifting.

\noindent \textbf{Pairwise alignment.}
In our IRLLRec, textual intents and interaction intents are treated as two distinct views. Contrastive Learning (CL) achieves this alignment strategy by maximizing representation consistency between positive samples in both views while pushing negative samples apart in the embedding space. For optimizing the representations of the two types of intent embeddings, the proposed objective function is as follows:
\begin{equation}
    \mathcal{L}_{\text{pair}} = \frac{1}{|\mathcal{B}|} \sum_{i \in \mathcal{B}} -\log \frac{\exp \left( s\left(\mathbf{r}_i, \mathbf{z}_i \right) / \tau \right)}{\sum_{j \in \mathcal{B}} \exp \left( s\left(\mathbf{r}_i, \mathbf{z}_j \right) / \tau \right)},
    \label{eq13}
\end{equation}
In a batch $\mathcal{B}$, the base model includes embeddings for three types of samples: user, positive, and negative. Therefore, we contrast them separately to obtain the losses $\mathcal{L}^{user}_{\text{pair}}$, $\mathcal{L}^{pos}_{\text{pair}}$ and $\mathcal{L}^{neg}_{\text{pair}}$.

\noindent \textbf{Translation alignment.}
Given the difficulty in obtaining positive and negative samples in interactions, we employ a positive pair alignment strategy to mitigate this limitation (e.g., only the intents of the same user or item are regarded as positive pairs). To better adapt to the inherent perturbations in input features, we add Gaussian noise \citep{2020gaussian} to both embeddings to enhance robustness:
\begin{equation}
\begin{aligned}
    \mathbf{r}' &= \mathbf{r} + \boldsymbol{\epsilon}_r \odot \mathbf{r}, \quad \boldsymbol{\epsilon}_r \sim \mathcal{N}(0, \mathbf{I}), \\
    \mathbf{z}' &= \mathbf{z} + \boldsymbol{\epsilon}_z \odot \mathbf{z}, \quad \boldsymbol{\epsilon}_z \sim \mathcal{N}(0, \mathbf{I}),
    \label{eq14}
\end{aligned}
\end{equation}
where $\boldsymbol{\epsilon}_r$ and $\boldsymbol{\epsilon}_r$ are auxiliary noise variables sampled from $\mathcal{N}(0, \mathbf{I})$, and each element in $\mathcal{N}(0, \mathbf{I})$ follows a standard normal distribution. As shown in the Figure \ref{fig:model}, we construct the diagonal elements as pairs of positive samples in set $\mathbb{R}$, while the other elements constitute negative samples in set $\mathbb{N}$. Therefore, in the text view, the contrastive loss is defined as follows:
\begin{equation}
    \mathcal{L}^{text}_{\text{tran}} = -\log \frac{\sum_{i \in \mathbb{R}} \exp\left( \eta \cdot \mathbf{r'}_i \cdot \mathbf{z'}_i^{\top} \right)}{\sum_{j \in \mathbb{R} \cup \mathbb{N}} \exp\left( \eta \cdot \mathbf{r'}_i \cdot \mathbf{z'}_j^{\top} \right)}
    \label{eq15}
\end{equation}
the $\eta$ represents the scaling factor, used to adjust the similarity intensity during the alignment process. $\mathcal{L}^{text}_{\text{tran}}$ is represents the text-interaction loss. Similar to the text view, the interaction view also has a loss $\mathcal{L}^{inter}_{\text{tran}}$, derived from the transposed similarity matrix. Finally, the overall loss is as follows:
\begin{equation}
    \mathcal{L}_{\text{IA}} = \lambda_1 (\mathcal{L}^{user}_{\text{pair}} + \mathcal{L}^{pos}_{\text{pair}} + \mathcal{L}^{neg}_{\text{pair}}) + \lambda_2 (\mathcal{L}^{text}_{\text{tran}} + \mathcal{L}^{inter}_{\text{tran}}).
    \label{eq16}
\end{equation}
where $\lambda_1$ and $\lambda_2$ represent the adjustable weights for two alignment strategies, respectively.

\subsection{Interaction-text Matching}
The information about interaction intents is often contained in certain parts of the text, so they tend to be noisy. The alignment process is typically weakly correlated, as the text may contain words unrelated to the true intents driving the interactions or intents not reflected in the interactions. To address this issue, we propose Interaction-text Matching (ITM), which is used to determine whether a pair of intents is matched or unmatched. ITM is a momentum model, serving as an evolving teacher model. 

Inspired by \citep{2020moco}, we maintain two momentum unimodal encoder $g'_t$ ($g_t$ is the text encoder) and $g'_i$ ($g_i$ is the graph encoder) for obtaining pseudo-labels for the teacher model. Specifically, we use $g'_t$ and $g'_i$ to calculate the representation similarities $s'(T, I) = g'_t(\mathbf{x})^\top g'_i(\mathbf{e})$ and $s'(I, T) = s'(T, I)^\top$, where $T$ and $I$ denote the textual and interaction perspectives, respectively. Next, we represent the student model's results $s(T, I) = \mathbf{z}^\top \mathbf{r}$ and $s(I, T) = s(T, I)^\top$ as $\mathbf{p}^{t2i}$ and $\mathbf{p}^{i2t}$, separately. Similarly, $s'(T, I)$ and $s'(I, T)$ denote $\mathbf{q}^{t2i}$ and $\mathbf{q}^{i2t}$. Formally, the ITM loss is defined as:
\begin{equation}
\begin{split}
    \mathcal{L}_{\text{ITM}} = &\ (1 - \alpha) \mathcal{L}_{\text{tran}} + \alpha \mathbb{E}_{(I, T) \sim \mathcal{D}} \Big[ 
    \text{KL}\left(\mathbf{q}^{i2t}(I) \parallel \mathbf{p}^{i2t}(I) \right) \\
    &\quad + \text{KL}\left(\mathbf{q}^{t2i}(T) \parallel \mathbf{p}^{t2i}(T) \right)\Big],
\end{split}
\label{eq17}
\end{equation}
where $\alpha$ is the balancing coefficient between the teacher and student models, and $i2t$ represents the interaction-to-text matching task, while $t2i$ is similar. To simplify, we test and set the weight is set to 0.4. $\text{KL}(\parallel)$ is employed to quantify the difference between two distributions. After each training iteration, the momentum model's parameters are updated using Exponential Moving Average (EMA) \citep{2021ALBEF}, smoothly transferring the student model's parameters to the teacher model to enhance the teacher model's stability:
\begin{equation}
    \theta_{\text{teacher}} = \beta \cdot \theta_{\text{teacher}} + (1 - \beta) \cdot \theta_{\text{student}}.
    \label{eq18}
\end{equation}
where $\beta$ is the momentum coefficient (e.g., 0.999). The momentum model is typically used to generate pseudo-labels, filtering out high-frequency noise in the student model and allowing the momentum model to capture more stable features.

\subsection{Model Analysis}
\noindent \textbf{Multi-task Training.}
Finally, we employ multi-task joint learning to complete the framework's optimization process, and the optimization objective of IRLLRec is as follows:
\begin{equation}
    \mathcal{L}_{\text{IRLLRec}} = \mathcal{L}_{\text{IA}} + \lambda_3 \mathcal{L}_{\text{ITM}} + \|\Theta\|_{2}^{2}.
    \label{eq19}
\end{equation}
where $\|\Theta\|_{2}^{2}$ are trainable model parameters and $L_2$ regularization.
% \noindent \textbf{Theoretical Analysis.}

\noindent \textbf{Time Complexity Analysis.}
We analyze the batch time complexity for this model-agnostic framework, excluding graph convolution, which is common to all base models. Let $d_t$ denote the dimension of the text representation and $d$ the dimension of user and item embeddings. The time complexity of the text encoder MLP encoding is $\mathcal{O}((M+N) d_t d)$, where $M$ and $N$ represent the number of users and items, respectively. Next, interaction intent encoder has a complexity of $\mathcal{O}((M+N) L K d)$, where $K$ is the number of latent intents and $L$ is the number of encoder layers. Additionally, the intent interaction reconstruction complexity is $\mathcal{O}(|\mathcal{G}| L d)$, which is used to remove noise from the intent. Finally, the complexity of the loss functions for IA and ITM are $\mathcal{O}(2 \mathcal{B} d + (M+N)^2 d)$ and $\mathcal{O}(M^2 + N^2)$, respectively, where $\mathcal{B}$ is a sampled batch. Momentum distillation doubles the complexity of the encoding process.

\begin{table}[h]
\captionsetup{justification=centering}
\caption{Statistics of the experimental datasets.}
\begin{adjustbox}{width=0.48\textwidth}
\begin{NiceTabular}{ccccc}
\toprule
\textbf{Dataset} & \textbf{\#Users} & \textbf{\#Items} & \textbf{\#Interactions} & \textbf{Density} \\
\midrule
Amazon-book  & 11,000 & 9,332  & 200,860 & $2.0 \times 10^{-3}$ \\
Yelp         & 11,091 & 11,010 & 277,535 & $2.3 \times 10^{-3}$ \\
Amazon-movie & 16,994 & 9,370  & 168,243 & $1.1 \times 10^{-3}$ \\
\bottomrule
\end{NiceTabular}
\end{adjustbox}
\label{table:dataset}
\end{table}

\begin{table*}[t]
\captionsetup{justification=centering}
\caption{Recommendation performance Imprvement of all backbone methods on different datasets in terms of Recall and NDCG. The superscript * indicates the Imprvement is statistically significant where the p-value is less than 0.05.}
\begin{adjustbox}{width=\textwidth}
\begin{NiceTabular}{c|cccccc|cccccc|cccccc}
\toprule[1pt]
\textbf{Dataset} & \multicolumn{6}{c}{\textbf{Amazon-book}} & \multicolumn{6}{c}{\textbf{Yelp}} & \multicolumn{6}{c}{\textbf{Amazon-movie}} \\
\midrule
\textbf{Metric} & \multicolumn{3}{c}{Recall} & \multicolumn{3}{c}{NDCG} & \multicolumn{3}{c}{Recall} & \multicolumn{3}{c}{NDCG} & \multicolumn{3}{c}{Recall} & \multicolumn{3}{c}{NDCG} \\
\midrule
\textbf{Model} & @5 & @10 & @20 & @5 & @10 & @20 & @5 & @10 & @20 & @5 & @10 & @20 & @5 & @10 & @20 & @5 & @10 & @20 \\
\midrule
Semantic Only & 0.0081 & 0.0125 & 0.0199 & 0.0072 & 0.0088 & 0.0112 & 0.0013 & 0.0022 & 0.0047 & 0.0014 & 0.0018 & 0.0026 & 0.0029 & 0.0084 & 0.0117 & 0.0022 & 0.0069 & 0.0098 \\
AlphaRec & 0.0598 & 0.0941 & 0.1412 & 0.0605 & 0.0721 & 0.0873 & 0.0431 & 0.0726 & 0.1212 & 0.0493 & 0.0586 & 0.0752 & 0.0909 & 0.1213 & 0.1661 & 0.0706 & 0.0827 & 0.0958 \\
\midrule
LightGCN & 0.0570 & 0.0915 & 0.1411 & 0.0574 & 0.0694 & 0.0856 & 0.0421 & 0.0706 & 0.1157 & 0.0491 & 0.0580 & 0.0733 & 0.0796 & 0.1193 & 0.1661 & 0.0623 & 0.0761 & 0.0891 \\
KAR & 0.0596 & 0.0934 & 0.1416 & 0.0590 & 0.0705 & 0.0860 & 0.0437 & 0.0740 & 0.1194 & 0.0506 & 0.0602 & 0.0756 & 0.0824 & 0.1195 & 0.1653 & 0.0645 & 0.0775 & 0.0905 \\
RLMRec-Con & \underline{0.0608} & \underline{0.0969} & \underline{0.1483} & \underline{0.0606} & \underline{0.0734} & \underline{0.0903} & \underline{0.0445} & \underline{0.0754} & \underline{0.1230} & \underline{0.0518} & \underline{0.0614} & \underline{0.0776} & \underline{0.0834} & \underline{0.1193} & 0.1697 & \underline{0.0652} & \underline{0.0778} & 0.0918 \\
RLMRec-Gen & 0.0596 & 0.0948 & 0.1446 & 0.0605 & 0.0724 & 0.0887 & 0.0435 & 0.0734 & 0.1209 & 0.0505 & 0.0600 & 0.0761 & 0.0823 & 0.1185 & \underline{0.1705} & 0.0643 & 0.0770 & \underline{0.0921} \\
IRLLRec & \textbf{0.0643*} & \textbf{0.1009*} & \textbf{0.1538*} & \textbf{0.0638*} & \textbf{0.0765*} & \textbf{0.0938*} & \textbf{0.0460*} & \textbf{0.0781*} & \textbf{0.1278*} & \textbf{0.0542*} & \textbf{0.0642*} & \textbf{0.0810*} & \textbf{0.0950*} & \textbf{0.1341*} & \textbf{0.1855*} & \textbf{0.0752*} & \textbf{0.0888*} & \textbf{0.1032*} \\
Improv. & 5.72\% & 4.13\% & 3.71\% & 5.21\% & 4.28\% & 3.85\% & 3.39\% & 3.63\% & 3.90\% & 4.57\% & 4.55\% & 4.39\% & 13.96\% & 12.40\% & 8.82\% & 15.37\% & 14.11\% & 12.03\% \\
\midrule
SGL & 0.0637 & 0.0994 & 0.1473 & 0.0632 & 0.0756 & 0.0913 & 0.0432 & 0.0722 & 0.1197 & 0.0501 & 0.0592 & 0.0753 & 0.0917 & 0.1239 & 0.1658 & 0.0722 & 0.0835 & 0.0953 \\
KAR & 0.0595 & 0.0941 & 0.1435 & 0.0596 & 0.0713 & 0.0875 & 0.0441 & 0.0735 & 0.1200 & 0.0510 & 0.0599 & 0.0757 & 0.0889 & 0.1230 & 0.1647 & 0.0714 & 0.0838 & 0.0958 \\
RLMRec-Con & \underline{0.0655} & \underline{0.1017} & 0.1528 & \underline{0.0652} & \underline{0.0778} & 0.0945 & 0.0452 & 0.0763 & 0.1248 & 0.0530 & 0.0626 & 0.0790 & \underline{0.0916} & 0.1275 & 0.1747 & \underline{0.0722} & \underline{0.0847} & \underline{0.0978} \\
RLMRec-Gen & 0.0644 & 0.1015 & \underline{0.1537} & 0.0648 & 0.0777 & \underline{0.0947} & \textbf{0.0467} & \underline{0.0771} & \underline{0.1263} & \underline{0.0537} & \underline{0.0631} & \underline{0.0798} & 0.0892 & \underline{0.1292} & \underline{0.1760} & 0.0701 & 0.0841 & 0.0972 \\
IRLLRec & \textbf{0.0671*} & \textbf{0.1031*} & \textbf{0.1539*} & \textbf{0.0680*} & \textbf{0.0803*} & \textbf{0.0964*} & \underline{0.0465} & \textbf{0.0784*} & \textbf{0.1275*} & \textbf{0.0541} & \textbf{0.0640*} & \textbf{0.0805} & \textbf{0.0963*} & \textbf{0.1349*} & \textbf{0.1857*} & \textbf{0.0757*} & \textbf{0.0892*} & \textbf{0.1035*} \\
Improv. & 2.50\% & 1.40\% & 0.14\% & 4.33\% & 3.16\% & 1.75\% & -0.43\% & 1.66\% & 0.95\% & 0.74\% & 1.36\% & 0.88\% & 5.15\% & 4.46\% & 5.52\% & 4.86\% & 5.42\% & 5.82\% \\
\midrule
SimGCL & 0.0618 & 0.0992 & 0.1512 & 0.0619 & 0.0749 & 0.0919 & 0.0467 & 0.0772 & 0.1254 & 0.0546 & 0.0638 & 0.0801 & 0.0925 & 0.1276 & 0.1700 & 0.0738 & 0.0861 & 0.0978 \\
KAR & 0.0623 & 0.1003 & 0.1520 & 0.0633 & 0.0756 & 0.0924 & 0.0466 & 0.0768 & 0.1265 & 0.0538 & 0.0636 & 0.0804 & 0.0930 & 0.1285 & 0.1685 & 0.0754 & 0.0876 & 0.1005 \\
RLMRec-Con & \underline{0.0633} & \underline{0.1011} & \underline{0.1552} & \underline{0.0633} & \underline{0.0765} & \underline{0.0942} & \underline{0.0470} & \underline{0.0784} & \underline{0.1292} & \underline{0.0546} & \underline{0.0642} & \underline{0.0814} & \underline{0.0979} & 0.1323 & 0.1763 & 0.0775 & 0.0896 & 0.1020 \\
RLMRec-Gen & 0.0617 & 0.0991 & 0.1525 & 0.0622 & 0.0752 & 0.0925 & 0.0464 & 0.0767 & 0.1267 & 0.0541 & 0.0634 & 0.0803 & 0.0969 & \underline{0.1324} & \underline{0.1781} & \underline{0.0780} & \underline{0.0902} & \underline{0.1034} \\
IRLLRec & \textbf{0.0669*} & \textbf{0.1057*} & \textbf{0.1575*} & \textbf{0.0666*} & \textbf{0.0802*} & \textbf{0.0970*} & \textbf{0.0493*} & \textbf{0.0818*} & \textbf{0.1328*} & \textbf{0.0571*} & \textbf{0.0671*} & \textbf{0.0844*} & \textbf{0.1011*} & \textbf{0.1374*} & \textbf{0.1863*} & \textbf{0.0817*} & \textbf{0.0946*} & \textbf{0.1082*} \\
Improv. & 5.66\% & 4.51\% & 1.46\% & 5.21\% & 4.84\% & 2.95\% & 4.81\% & 4.31\% & 2.77\% & 4.54\% & 4.52\% & 3.71\% & 3.21\% & 3.81\% & 4.63\% & 4.72\% & 4.83\% & 4.62\% \\
\midrule
DCCF & 0.0662 & 0.1019 & 0.1517 & 0.0658 & 0.0780 & 0.0943 & 0.0468 & 0.0778 & 0.1249 & 0.0543 & 0.0640 & 0.0800 & 0.0965 & 0.1317 & 0.1723 & 0.0763 & 0.0905 & 0.1014 \\
KAR & 0.0665 & 0.1022 & 0.1521 & 0.0657 & 0.0785 & 0.0949 & 0.0465 & 0.0784 & 0.1257 & 0.0539 & 0.0642 & 0.0808 & 0.0977 & 0.1334 & 0.1750 & 0.0778 & 0.0914 & 0.1029 \\
RLMRec-Con & 0.0665 & 0.1040 & \underline{0.1563} & 0.0668 & 0.0798 & 0.0968 & \underline{0.0486} & \underline{0.0813} & \underline{0.1321} & \underline{0.0561} & \underline{0.0663} & \underline{0.0836} & \underline{0.0987} & \underline{0.1349} & \underline{0.1790} & \underline{0.0796} & \underline{0.0921} & \underline{0.1048} \\
RLMRec-Gen & \underline{0.0666} & \underline{0.1046} & 0.1559 & \underline{0.0670} & \underline{0.0801} & \underline{0.0969} & 0.0475 & 0.0785 & 0.1281 & 0.0549 & 0.0646 & 0.0815 & 0.0980 & 0.1349 & 0.1778 & 0.0792 & 0.0918 & 0.1041 \\
IRLLRec & \textbf{0.0702*} & \textbf{0.1076*} & \textbf{0.1590*} & \textbf{0.0690*} & \textbf{0.0822*} & \textbf{0.0992*} & \textbf{0.0506*} & \textbf{0.0843*} & \textbf{0.1360*} & \textbf{0.0581*} & \textbf{0.0689*} & \textbf{0.0864*} & \textbf{0.1012*} & \textbf{0.1374*} & \textbf{0.1826*} & \textbf{0.0812*} & \textbf{0.0939*} & \textbf{0.1066*} \\
Improv. & 5.41\% & 2.87\% & 1.73\% & 2.96\% & 2.60\% & 2.41\% & 4.04\% & 3.69\% & 2.96\% & 3.56\% & 3.91\% & 3.35\% & 2.53\% & 1.82\% & 2.00\% & 2.06\% & 1.98\% & 1.72\% \\
\midrule
BIGCF & 0.0662 & 0.1028 & 0.1552 & 0.0658 & 0.0784 & 0.0955 & 0.0458 & 0.0758 & 0.1237 & 0.0536 & 0.0627 & 0.0789 & 0.0947 & 0.1330 & 0.1793 & 0.0759 & 0.0893 & 0.1022 \\
KAR & 0.0670 & 0.1034 & 0.1559 & 0.0664 & 0.0792 & 0.0961 & 0.0446 & 0.0763 & 0.1245 & 0.0533 & 0.0630 & 0.0799 & 0.0952 & 0.1342 & 0.1808 & 0.0766 & 0.0904 & 0.1030 \\
RLMRec-Con & \underline{0.0685} & \underline{0.1052} & \underline{0.1574} & \underline{0.0680} & \underline{0.0807} & \underline{0.0977} & \underline{0.0470} & \underline{0.0784} & \underline{0.1284} & \underline{0.0548} & \underline{0.0646} & \underline{0.0815} & \underline{0.0987} & \underline{0.1373} & \underline{0.1845} & \underline{0.0779} & \underline{0.0914} & \underline{0.1046} \\
RLMRec-Gen & 0.0676 & 0.1044 & 0.1555 & 0.0654 & 0.0796 & 0.0965 & 0.0467 & 0.0764 & 0.1259 & 0.0545 & 0.0635 & 0.0802 & 0.0954 & 0.1339 & 0.1814 & 0.0765 & 0.0899 & 0.1031 \\
IRLLRec & \textbf{0.0693*} & \textbf{0.1082*} & \textbf{0.1631*} & \textbf{0.0694*} & \textbf{0.0831*} & \textbf{0.1010*} & \textbf{0.0495*} & \textbf{0.0823*} & \textbf{0.1342*} & \textbf{0.0576*} & \textbf{0.0678*} & \textbf{0.0854*} & \textbf{0.1048*} & \textbf{0.1431*} & \textbf{0.1916*} & \textbf{0.0845*} & \textbf{0.0979*} & \textbf{0.1115*} \\
Improv. & 1.23\% & 2.83\% & 3.63\% & 2.09\% & 2.90\% & 3.32\% & 5.32\% & 4.97\% & 4.52\% & 5.11\% & 4.95\% & 4.79\% & 6.16\% & 4.20\% & 3.87\% & 8.44\% & 7.11\% & 6.62\% \\
\bottomrule
\end{NiceTabular}
\end{adjustbox}
\label{table:result}
\end{table*}

\section{EXPERIMENTS}
In this section, we conduct extensive experiments and answer the following research questions:
\begin{itemize}[leftmargin=*]
\item \textbf{RQ1:} How does IRLLRec compare to the current state-of-the-art (SOTA) framework in terms of performance?
\item \textbf{RQ2:} Are the key components in our IRLLRec delivering the expected performance gains?
\item \textbf{RQ3:} What are the reasons for IRLLRec's superior performance?
\item \textbf{RQ4:} How do different hyperparameters affect IRLLRec?
\end{itemize}

\subsection{Experiment Setup}
\subsubsection{\textbf{Datasets}}
The performance of IRLLRec has been validated on three real-world datasets. A statistical overview of all the datasets is presented in Table \ref{table:dataset}. 
\textbf{Amazon-book}\citep{2024rlmrec} contains users' book purchase, rating, and review records from Amazon.
\textbf{Yelp}\footnote{https://business.yelp.com/data/resources/open-dataset/} covers users' reviews and rating information for local businesses.
\textbf{Amazon-movie}\footnote{https://amazon-reviews-2023.github.io/} provides users' movie-watching records, ratings, and review data from Amazon.
The data processing follows the previous work \citep{2024rlmrec, 2017ncf}, where interactions with ratings below 2 in Amazon-book, below 3 in Yelp, and below 3 in Amazon-movie are filtered out. Next, we perform k-core filtering. Amazon-book and Yelp datasets are split into training, validation, and test sets in a 3:1:1 ratio, while Amazon-movie is split in a ratio of 8:1:1.

\subsubsection{\textbf{Baselines and Base Models}}
Our IRLLRec employs an intent alignment component to enhance representation-based recommenders. Therefore, we selected representative models as base models. To evaluate the effectiveness of the method, we selected two state-of-the-art methods as baselines.

\noindent \textbf{Baselines.}
% \textbf{KAR} \citep{2024KAR} introduces external knowledge of user preferences and factual knowledge about items, which are compressed into vectors to improve recommendations. \textbf{RLMRec} \citep{2024rlmrec} proposes a framework that leverages the representation learning empowered by LLMs and designs contrastive and generative alignment methods to enhance recommendations. \textbf{AlphaRec} \citep{2024alpharec} replaces ID-based embeddings with language embeddings and combines GCN and CL to achieve a simple yet effective recommendation paradigm.
\begin{itemize}[leftmargin=*]
\item \textbf{KAR} \citep{2024KAR} introduces external knowledge of user preferences and factual knowledge about items, which are compressed into vectors to improve recommendations.
\item \textbf{RLMRec} \citep{2024rlmrec} proposes a framework that leverages the representation learning empowered by LLMs and designs contrastive and generative alignment methods to enhance recommendations.
\item \textbf{AlphaRec} \citep{2024alpharec} replaces ID-based embeddings with language embeddings and combines GCN and CL to achieve a simple yet effective recommendation paradigm.
\end{itemize}

\noindent \textbf{Base Models.}
% \textbf{LightGCN} \citep{2020lightgcn} removes the feature transformation and nonlinear activations of GCN to achieve light recommendation.
% \textbf{SGL} \citep{2021sgl} generates contrast views by edge dropout to aid contrastive learning to enhance recommendation.
% \textbf{SimGCL} \citep{2022simgcl} considers the relationship between neighbor nodes to enhance collaborative filtering.
% \textbf{DCCF} \citep{2023dccf} enhances self-supervised signals by learning disentangled representations with global context.
% \textbf{BIGCF} \citep{2024bigcf} explores the individuality and collectivity of intents behind interactions for collaborative filtering.
\begin{itemize}[leftmargin=*]
\item \textbf{LightGCN} \citep{2020lightgcn} removes the feature transformation and nonlinear activations of GCN to achieve light recommendation.
\item \textbf{SGL} \citep{2021sgl} generates contrast views by edge dropout to aid contrastive learning to enhance recommendation.
\item \textbf{SimGCL} \citep{2022simgcl} considers the relationship between neighbor nodes to enhance collaborative filtering.
\item \textbf{DCCF} \citep{2023dccf} enhances self-supervised signals by learning disentangled representations with global context.
\item \textbf{BIGCF} \citep{2024bigcf} explores the individuality and collectivity of intents behind interactions for collaborative filtering.
\end{itemize}

\begin{figure}[t]
    \centering
    \includegraphics[width=\linewidth]{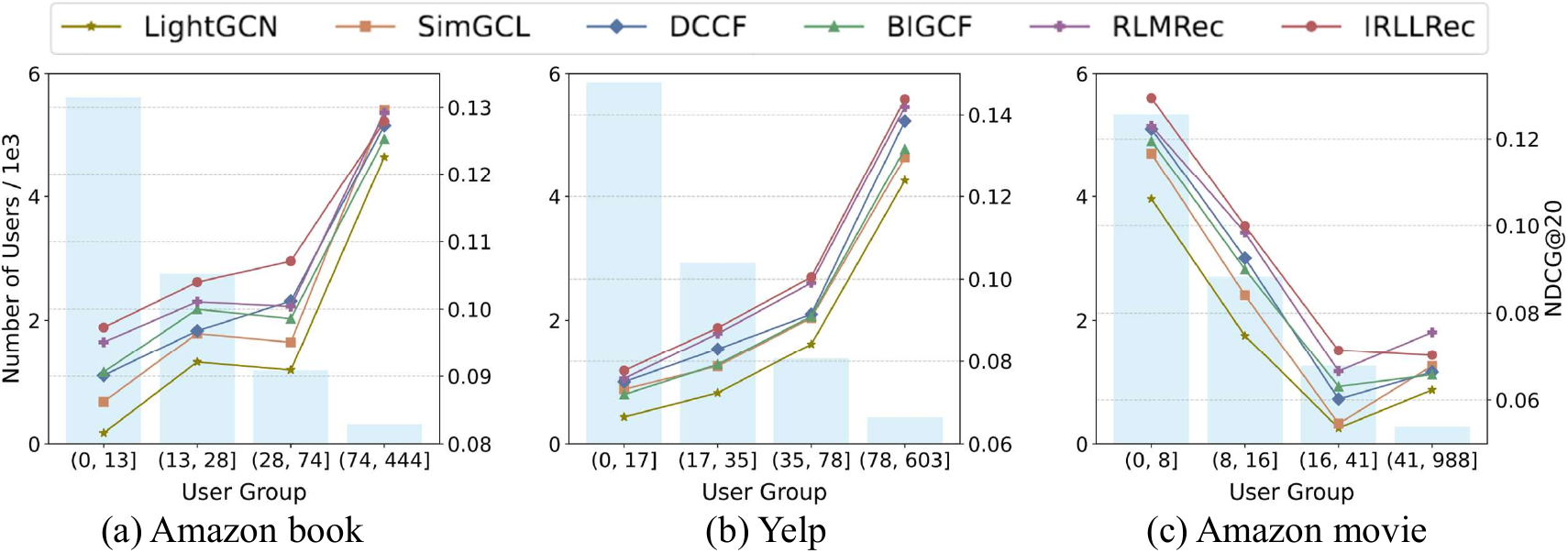}
    \caption{Performance comparison of different sparsity levels. \textnormal{The bar graph shows users' number per group on the left y-axis, and the line graph shows the performance of each method w.r.t. NDCG@20 on the right y-axis.}}
    \label{fig:sparsity}
\end{figure}

\begin{figure}[ht]
    \centering
    \includegraphics[width=\linewidth]{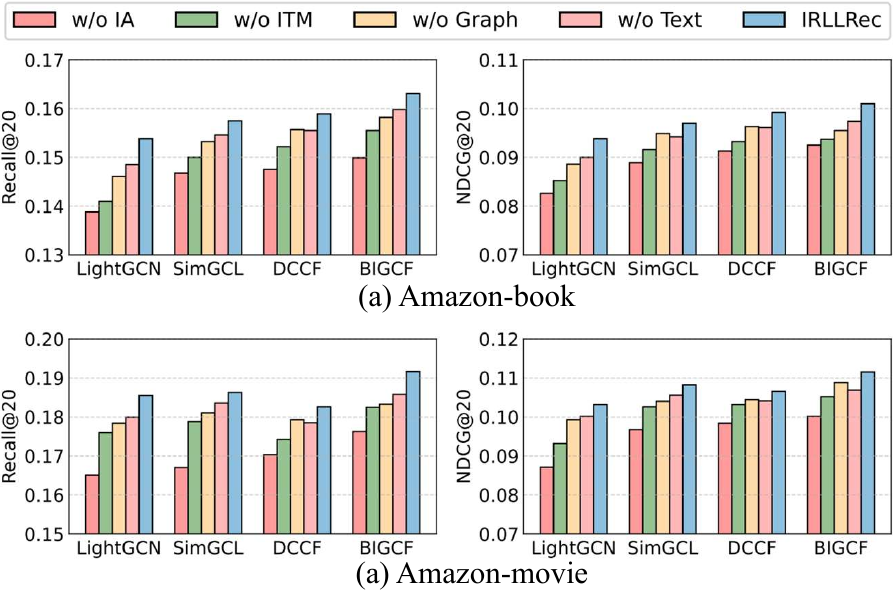}
    \caption{Ablation studies of model variants on the Amazon book and movie datasets w.r.t. Recall@20 and NDCG@20.}
    \label{fig:ablation}
\end{figure}

\subsubsection{\textbf{Implementations}}
To ensure fair comparison, we adopted the sampling method and dataset format from the open-source framework SSLRec \citep{2024sslrec}. In the experimental setup, the embedding dimension of the models and the transformed LLM embeddings were uniformly set to 32. All models, including baselines and base models, were trained using the Adam optimizer with Xavier \citep{2010xavier} initialization for embeddings, a fixed batch size of 4096, and a learning rate of 1e-3. Additionally, early stopping was applied based on the model's performance on the validation set to prevent overfitting. For textual intent generation, we used OpenAI's GPT-4o-mini for intent generation and text-embeddings-3-large \citep{2022textemb} for semantic representation to ensure accuracy in intent expression and consistency in semantic representation.

\subsection{Performance Comparisons(RQ1)}
\subsubsection{\textbf{Comparisons w.r.t. Overall Performance.}}
Table \ref{table:result} presents the improvement in recommendation performance of the intent-based, model-agnostic framework across three public datasets. We will discuss notable phenomena and provide potential explanations. The experimental results are averaged over 5 runs.
\begin{itemize}[leftmargin=*]
    \item Compared to all baselines, IRLLRec shows significant improvements across all metrics (5.72\% and 15.37\% for Amazon book and movie, respectively, and 5.32\% for Yelp). The only exception is the recall@5 metric on Yelp, which is 0.43\% lower than RLMRec-Gen. We believe this is due to the random perturbation of the graph structure in SGL \citep{2021sgl}, which disrupts intent propagation and shifts the alignment direction towards noise, yet still highlights the framework's superiority and rationale.
    \item Both KAR \citep{2024KAR} and RLMRec \citep{2024rlmrec} are LLM-enhanced methods that improve base models by generating textual user/item descriptions. However, KAR is less stable than RLMRec due to its lack of integration between textual knowledge and user behavior, making it more susceptible to noise. Our IRLLRec addresses this limitation effectively through intent alignment and enhance recommendations by aligning fine-grained multimodal intents.
    \item The AlphaRec's poor performance may stem from our adherence to fairness, where we use RLMRec’s profile representation as the item embedding instead of the proposed item title.
\end{itemize}

\subsubsection{\textbf{Comparisons w.r.t. Data Sparsity}}
A common challenge in existing models is data sparsity, where users with few interactions dominate. As shown in Figure \ref{fig:sparsity}, we divide users into four groups based on interaction count across three datasets, ensuring roughly equal total interactions per group. For instance, in the Amazon book dataset, the first group includes 5612 users with no more than 13 interactions, meaning over 50\% of users in the test set have limited interaction records. We observe that as interaction scale increases, the model generally makes more accurate recommendations (i.e., higher NDCG value), suggesting that `cold-start' (not strictly defined) users are a key limitation in recommendations. Compared to RLMRec, IRLLRec improves performance in the sparsest user group by 2.31\%, 2.51\%, and 5.20\%, respectively, demonstrating that multimodal intent better captures the preferences of these ‘cold-start’ users, leading to more accurate recommendations.

\subsection{Ablation Study (RQ2)}
\subsubsection{\textbf{Impact of key components.}}
In this section, we assess the impact of key components in IRLLRec and provide potential explanations for the results. The following model variants are evaluated:
\begin{itemize}[leftmargin=*]
\item IRLLRec w/o IA: Removes the intent alignment module;
\item IRLLRec w/o ITM: Removes the interaction-text matching;
\item IRLLRec w/o Graph: Replaces GCN \citep{2020lightgcn} operation on interaction intent embeddings with user and item embeddings;
\item IRLLRec w/o Text: Replaces MLP mapping of text intent embeddings with direct alignment.
\end{itemize}
Figure \ref{fig:ablation} presents the experimental results on the Amazon book and movie datasets, leading to the following insights:

Experimental results demonstrate that IA contributes the most to recommendation performance, with its absence often leading to worse outcomes than the base model. This highlights that the a single modality's intent cannot capture fine-grained preferences; instead, the two intents must be brought closer together in their spatial distance while filtering noise \citep{2024DALR}. In contrast, the ITM module ranks second in importance, consistently outperforming the base model due to its momentum distillation mechanism, which effectively aligns the two types of intents.

\begin{figure}[t]
    \centering
    \includegraphics[width=\linewidth]{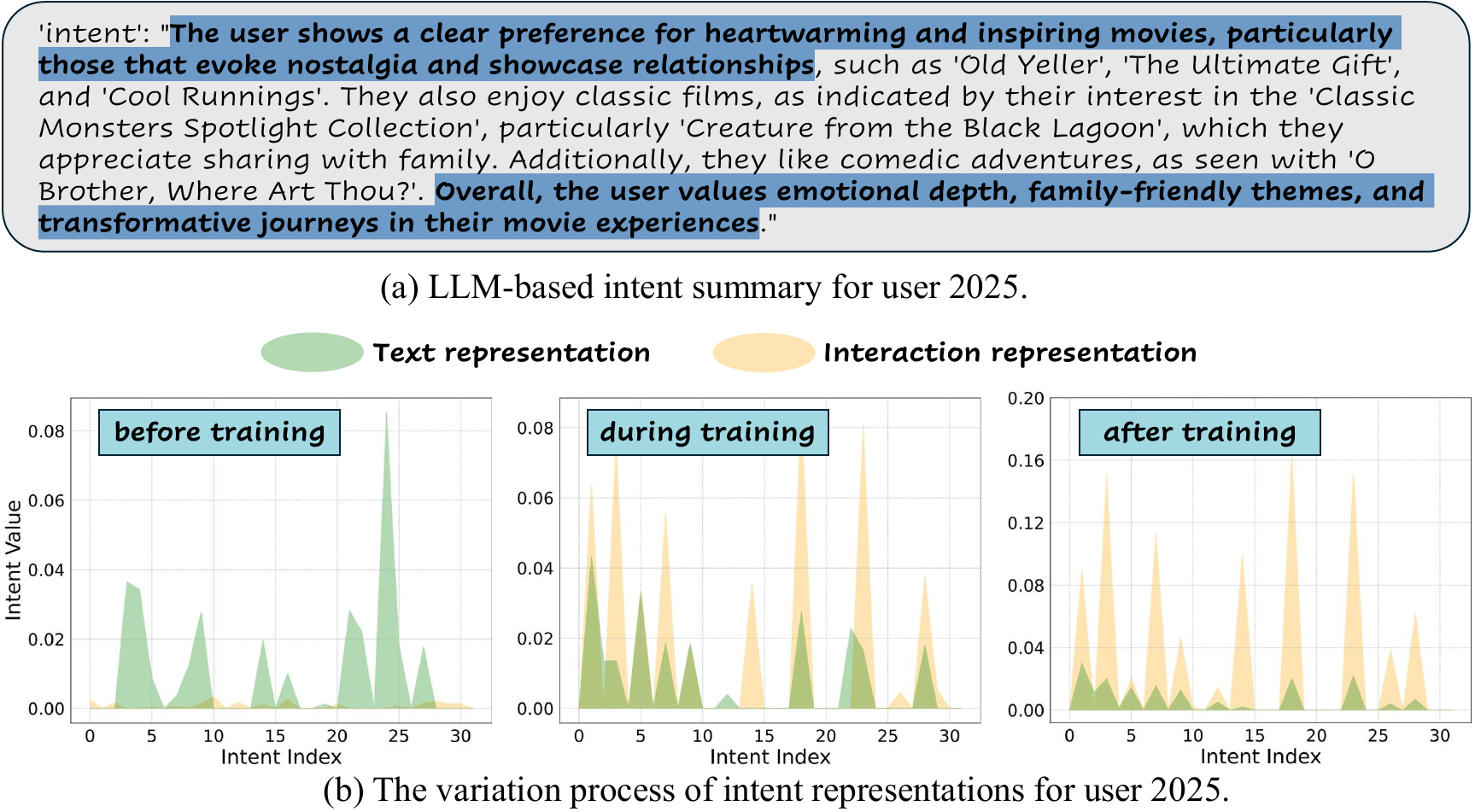}
    \caption{Case study on Amazon movie.}
    \label{fig:case_study}
\end{figure}

\begin{figure}[ht]
    \centering
    \includegraphics[width=\linewidth]{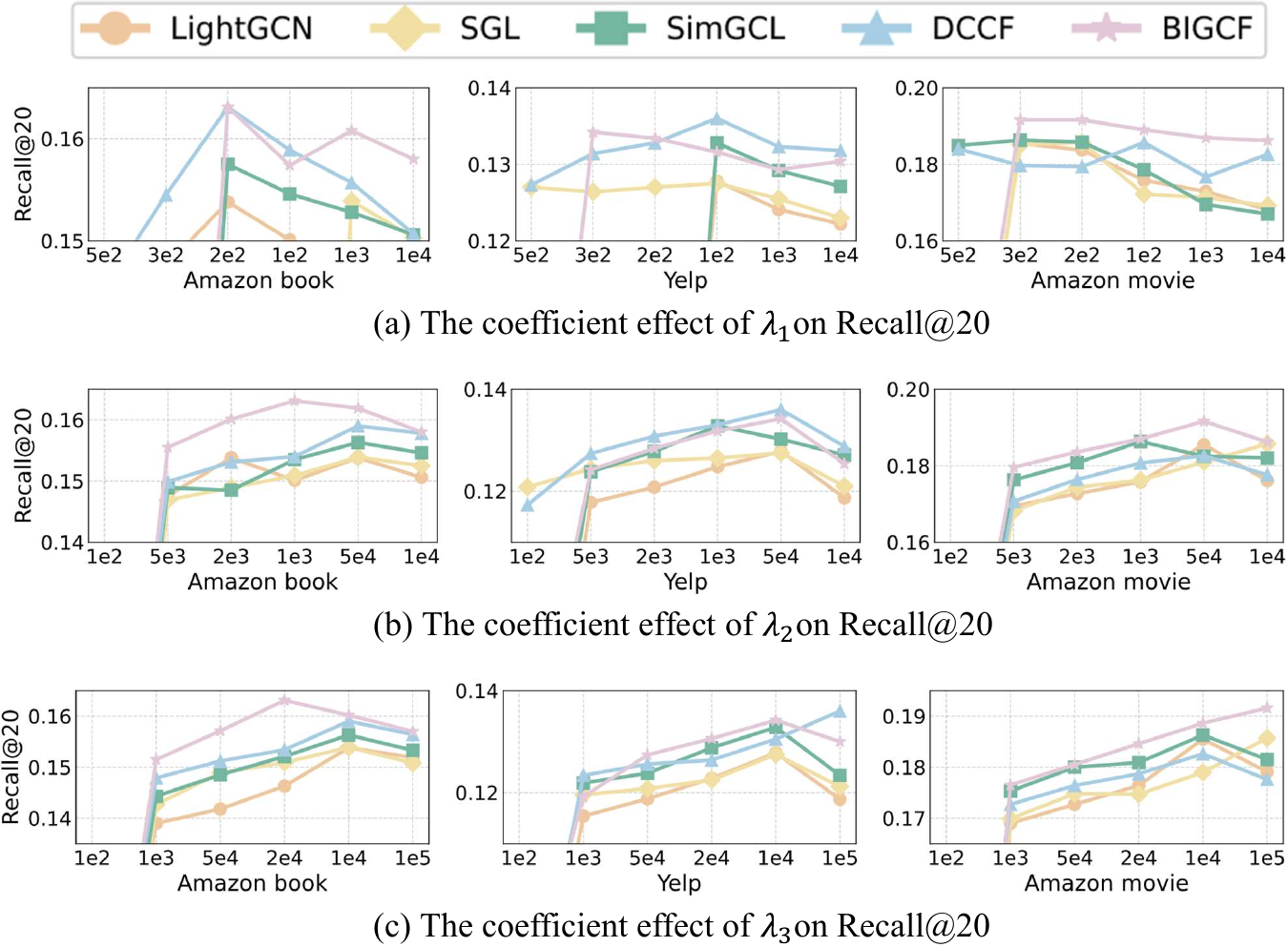}
    \caption{Performance comparison w.r.t. loss coefficients applied to the base model by IRLLRec.}
    \label{fig:hyper_param}
\end{figure}

Subsequently, we further decompose the framework into a text encoder and a graph encoder to evaluate the necessity of the dual-tower architecture. Notably, removing the graph encoder typically results in poorer performance. RLMRec seeks to align user representations with profile representations, highlighting that text-based intent representation goes beyond a simple user profile, capturing fine-grained preferences. Overall, substituting either encoder leads to a loss of IA's information, emphasizing the advantage of the dual-tower architecture.

\begin{table*}[t]
\captionsetup{justification=centering}
    \caption{Ablation studies of different semantic representations on Amazon-movie datasets w.r.t. Recall@20 and NDCG@20.}
    \begin{adjustbox}{width=0.95\textwidth}
    \begin{NiceTabular}{c|c|cc|cc|cc|cc|cc}
    \toprule[1pt] 
    \multirow{2}{*}{\textbf{Dataset}} & \multirow{2}{*}{\textbf{Variants}} & \multicolumn{2}{c}{\textbf{LightGCN}} & \multicolumn{2}{c}{\textbf{SGL}} & \multicolumn{2}{c}{\textbf{SimGCL}} & \multicolumn{2}{c}{\textbf{DCCF}} & \multicolumn{2}{c}{\textbf{BIGCF}} \\ 
    \cmidrule{3-12}
    &   & R@20  & N@20  & R@20  & N@20 & R@20  & N@20  & R@20  & N@20  & R@20  & N@20 \\
    \midrule
                    & Llama3-8B-Instruct      \citep{2024llama3}        & 0.1816  & 0.1005  & 0.1804  & 0.1003  & 0.1806  & 0.1036  & 0.1781  & 0.1025 & 0.1859  & 0.1074 \\
    \textbf{Amazon} & text-embedding-ada-002  \citep{2022textemb}       & 0.1837  & 0.1026  & 0.1835  & 0.1029  & 0.1844  & 0.1070  & 0.1805  & 0.1059 & 0.1898  & 0.1102 \\
    \textbf{Movie}  & text-embeddings-3-large \citep{2022textemb}       & \underline{0.1855}  & \underline{0.1032}  & \underline{0.1857}  & \underline{0.1035}  & \underline{0.1863}  &                                                                          \underline{0.1082}  & \underline{0.1826}  & \underline{0.1066}  & \underline{0.1916}  & \underline{0.1115} \\
                    & SFR-Embedding-Mistral   \citep{2024sfrembedding}  & \textbf{0.1864}  & \textbf{0.1040}  & \textbf{0.1866}  & \textbf{0.1043}  & \textbf{0.1874}  
                                                                        & \textbf{0.1084}  & \textbf{0.1832}  & \textbf{0.1076}  & \textbf{0.1935}  & \textbf{0.1123} \\ 
    \bottomrule[1pt]
    \end{NiceTabular}
    \end{adjustbox}
    \label{table:semantic}
\end{table*}

\subsubsection{\textbf{Impact of different semantic representations.}}
To assess the impact of text embedding models, we tested additional transformation models (e.g., Llama3 and GPT) to evaluate the effectiveness of LLMs for intent summarization. As shown in Table \ref{table:semantic}, despite Llama3’s larger embedding dimension (4096), its performance is the worst. This is due to the 8B model's insufficient parameter count, limiting its ability to capture fine-grained user preferences. In contrast, SFR-Embedding-Mistral, the latest model, achieved the best recommendation performance, demonstrating significant progress in LLM-based semantic transformation. Overall, both locally deployable and closed-source LMs significantly enhance the base model, highlighting the importance of intent summarization.

\subsection{Case Study (RQ3)}
We investigate the impact of interaction-text matching on intent alignment. Figure \ref{fig:case_study} (b) shows a case study where we split the 32-dimensional multimodal intent representation of user 2025 into individual values by index. Before training, the matching score (vector dot product) between the two representations was -0.0005, during training it increased to 0.0270, and after training, it reached 0.0399. Initially, the interaction-based intent representation was indistinguishable, but over time, it developed clearer preferences, and the bias in the multimodal intent representation decreased. Notably, the trained multimodal intent focuses on the head and tail, as shown in the blue sections of Figure \ref{fig:case_study} (a). This demonstrates that IRLLRec effectively aligns interaction-driven intents with LLM-summarized intents, validating Challenge 2.

\subsection{Hyperparameter Sensitivity (RQ4)}
In this study, we keep the parameters of RLMRec fixed while analyzing the loss coefficients of the IA and ITM components, corresponding to $\lambda_1$ and $\lambda_2$ in Eq. \ref{eq16}, and $\lambda_3$ in Eq. \ref{eq19}.
\begin{itemize}[leftmargin=*]
    \item As shown in Figure \ref{fig:hyper_param} (a), pairwise alignment in IA plays a key role, with optimal performance typically achieved between 0.01 and 0.03. Truncated cases in the figure are due to excessive non-primary losses that hinder model convergence and are not displayed. This suggests that alignment is the primary source of performance gains, with interaction-based alignment effectively integrating user behavior and semantic information.
    \item Translation alignment and interaction-text matching in IA exhibit clear trends in Figure \ref{fig:hyper_param} (b) and (c). For instance, recall typically peaks as $\lambda_2$ decreases, then gradually declines, with a similar pattern for $\lambda_3$. This indicates that these two components must maintain a delicate balance with pairwise alignment, where combining alignment and matching effectively addresses the noise issue in multimodal intents.
\end{itemize}

\section{RELATED WORK}
\subsection{LLMs for Recommendation}
Large language models (LLMs) have gained significant attention in recommender systems for their advanced language understanding and reasoning abilities. Research in this area primarily follows three paradigms: LLMs as recommenders, enhancers, and encoders. LLMs as recommenders \citep{2024llara, 2023llamarec}: These methods use user interaction histories as prompts to guide LLMs in selecting recommendation targets from a candidate set. TallRec \citep{2023tallrec} fine-tunes Llama on constructed recommendation datasets, enhancing LLM decision-making in recommendations. RosePO \citep{2024rosepo} employs smoothing personalized preference optimization to fine-tune LLMs, improving performance while ensuring the recommender remains "helpful and harmless". LLMs as enhancers \citep{2024DALR, 2023VQ-Rec}: RLMRec \citep{2024rlmrec} introduces a framework leveraging LLM-driven representation learning, with contrastive and generative alignment methods to improve recommendations. AlphaRec \citep{2024alpharec} replaces ID-based embeddings with language embeddings and combines GCN and CL for a simple yet effective recommendation approach. LLMs as encoders \citep{2023chatrec, 2024BLaIR}: EasyRec \citep{2024easyrec} leverages collaborative information and textual data from users and items to retrain language models for recommendation, achieving impressive performance in zero-shot scenarios. Despite their impact in respective fields, they overlook the potential of LLM-based intents to enhance interpretability.

\subsection{Disentanglement-based Recommendation}
Disentanglement-based methods generally focus on modeling user-item interactions by projecting them into distinct feature spaces \citep{2019MEIRec, 2018VAECF}. For instance, MacridVAE \citep{2019MacridVAE} leverages variational autoencoders to encode various user intents \citep{2019DGNN}. DGCF \citep{2020DGCF} employs graph neural networks to learn disentangled user representations. DisenHAN \citep{2020disenhan} utilizes meta-relation decomposition along with disentangled propagation layers to capture semantic meanings. In the case of CDR \citep{2021CDR}, a dynamic routing mechanism is developed to characterize the correlations among user intents for embedding denoising. KGIN \citep{2021kgin} introduces the concept of shared intents and uses an item-side knowledge graph to capture user’s path-based intents. Some innovative approaches have started integrating contrastive learning into intent modeling, such as ICLRec \citep{2022ICLRec}, DCCF \citep{2023dccf}, and BIGCF \citep{2024bigcf}. DCCF \citep{2023dccf} enhances self-supervised signals by learning disentangled representations with a global context, while BIGCF investigates the individuality and collectivity of intents behind interactions for collaborative filtering. While effective, multimodal intents present a promising avenue for exploration.

\section{CONCLUSION}
The paper leveraged LLMs to infer the semantic intents of users and items, and employed joint distribution and graph convolution to capture interaction intents. We proposed a model-agnostic framework, IRLLRec, incorporated a dual-tower model where textual and interaction intents were processed by separate encoders. We applied pairwise alignment to capture common features between the two intent types, reduced spatial differences, while translation alignment disrupted multimodal representations to enhance robustness against input noise. Additionally, we introduced an interaction-text matching method with momentum distillation for teacher-student learning of fused intent representations. Experiments on three public datasets validated the superiority of IRLLRec.

\begin{acks}
This work is supported by the National Science Foundation of China (No. 62206002, No. 62272001 and No. 62206004), Natural Science Foundation of Anhui Province (No. 2208085QF195) and Xunfei Zhiyuan Digital Transformation Innovation Research Special for Universities (No. 2023ZY001).
\end{acks}

\clearpage
\bibliographystyle{ACM-Reference-Format}
\balance
\bibliography{sample-base}

\end{document}